\RequirePackage[2020-02-02]{latexrelease}
\documentclass[aps,prd,groupedaddress,graphicx,nofootinbib]{revtex4}
\usepackage{amsmath,amssymb,graphics,graphicx,color,epsf}
\usepackage{subfigure}
\usepackage[scanall]{psfrag}  
\usepackage{tikz}

\begin{document}

\title{Just some simple (but nontrivial) analytical solutions for de Broglie-Bohm quantum cosmology}

\author{Chia-Min Lin}

\affiliation{Fundamental General Education Center, National Chin-Yi University of Technology, Taichung 41170, Taiwan}



\begin{abstract}

In this work, we provide some simple analytical solutions to the Wheeler-DeWitt equation for the minisuperspace applied to de Broglie-Bohmian quantum cosmology for particular potentials of a scalar matter field $\phi$. One solution describes a constantly rolling scalar field with a quadratic potential plus a negative cosmological constant. The scale factor is given by a Gaussian function. The other solution describes a flat Minkowski universe with the potential of a hilltop form. 

\end{abstract}
\maketitle
\large
\baselineskip 18pt
\section{Introduction}
\label{sec1}

Quantum cosmology \cite{DeWitt:1967yk} (see \cite{Halliwell:1989myn, Bojowald:2015iga, Linde:1990flp, Kolb:1990vq} for reviews and more references therein) is an ambitious project. It is assumed that the whole universe is quantum mechanical and is described by a wave function of the universe $\Psi$ . Due to the small value of Planck's constant, quantum phenomena usually manifest themselves in microscopic scales such as the sub-atomic scale. If so, probably it is not the best idea to apply quantum mechanics to the most macroscopic thing in the universe, namely the universe itself! However, the idea of cosmic inflation invites us to contemplate that the universe may be created as small as the Planck length via quantum fluctuations in a spacetime foam, and at least it may be described quantum mechanically before it inflates to become too big. Therefore the study of quantum cosmology usually concerns initial conditions in cosmology. The idea of cosmic inflation solves the initial condition problems of the hot big bang model \cite{Guth:1980zm}. One hope is that quantum cosmology may solve the initial condition problems of inflation. Surely there are initial conditions for inflation to happen or not to happen. On the other hand, is there any role played by the wave function of the universe after it becomes large? We may be able to address this question in this work.

In order to be more specific, one usually starts with a Hamiltonian formulation of general relativity through the Arnowitt, Deser, Minser (ADM) decomposition. 
\begin{equation}
ds^2=(Ndt)^2-h_{ij}(N^idt+dx^i)(N^jdt+dx^j),
\end{equation}
where $N$ is the lapse function, $N_i$ is the shift vector, and $h_{ij}$ is an induced spatial metric. The Wheeler-DeWitt equation even without a matter field is a complicated functional differential equation for the wave function\footnote{A more precise name in this section is to call $\Psi$ the wave functional.} $\Psi$:
\begin{equation}
\left[ \frac{G_{ijkl}}{(16 \pi G)^2}\frac{\delta}{\delta_{ij}} \frac{\delta}{\delta_{kl}}+\frac{\sqrt{h} R_3}{16\pi G}\right]\Psi[h_{ij}]=0,
\label{eq1}
\end{equation}
where
\begin{equation}
G_{ijkl}=\frac{1}{2}(h_{ik}h_{jl}+h_{il}h_{jk}-h_{ij}h_{kl})
\end{equation}
is the inverse of the DeWitt metric
\begin{equation}
G^{ijkl}=\frac{1}{2}(h^{ik}h^{jl}+h^{il}h^{jk}-h^{ij}h^{kl}).
\end{equation}
The wave function $\Psi$ is defined on a space, known as superspace\footnote{This has nothing to do with the superspace in supersymmetry.} of all possible $h_{ij}$ with diffeomorphisms factored out. 
In principle, there are many solutions of the Wheeler-DeWitt that depend on different initial or boundary conditions.
 
There are conceptual problems concerning the physical meaning of the wave function of the universe. There is the problem of time, which concerns the fact that the wave function of the universe does not depend (explicitly) on time. Because the Wheeler-DeWitt equation is of the zero-energy\footnote{It might be interesting to compare this with the argument given by Tryon in \cite{Tryon:1973xi} that according to the uncertainty principle $\Delta E \Delta t \sim \hbar$, if a universe is created via quantum fluctuation from the vacuum (or nothing) with a long lifetime $\Delta t \rightarrow \infty$, it must have net energy $\Delta E \rightarrow 0$.} form $H\Psi=0$. This reminds us of the quote allegedly from Einstein: \emph{time does not exist}. Yet time somehow feels quite real for us in the universe. There is also the measurement problem, which concerns how the wave function of the universe collapses according to the conventional Copenhagen interpretation \cite{Bohr:1935af}. Should there be an observer at the beginning of the universe when the universe was created? By definition, there is no observer outside the universe, no? In addition, there is also a problem about what is the physical meaning of the amplitude of the wave function. Is the square of it gives the possibility to create a universe? There are also technical problems, Eq.~(\ref{eq1}) is very difficult to solve. In practice, the infinite degrees of freedom of the superspace is restricted into a finite-dimensional subspace called minisuperspace, such as a homogeneous and isotropic universe. It would be good to find an analytical solution so that our physical intuition can work better to understand the physical meaning of what we found. If a solution is too complicated, one cannot see the forest for the trees. As long as the solution is nontrivial, probably the simplest one is the best.

Apparently, the subject of quantum cosmology is deeply connected to the foundation of quantum mechanics itself and also quantum gravity. There are open questions and the study of this subject requires a particularly open mind. In this work, we adopt the Bohmian interpretation \cite{Bohm:1951xw}, also known as the de Broglie-Bohm theory or pilot wave theory, which is briefly reviewed in the next section. 

\section{de Broglie-Bohm interpretation of quantum mechanics}
\label{sec2}
Let us consider the one-particle Schroedinger equation for a particle with mass $m$
\begin{equation}
i\frac{\partial \Psi}{\partial t}=-\frac{\nabla^2 \Psi}{2m}+V\Psi,
\end{equation}
where we have set $\hbar=1$. By substituting $\Psi=|\Psi|e^{iS}$ into the above equation, one of the relations we obtain is
\begin{equation}
\frac{\partial S}{\partial t}=-\left[ \frac{(\nabla S)^2}{2m}+V-\frac{\nabla^2 |\Psi|}{2m|\Psi|} \right].
\end{equation}
This can be interpreted as a Hamilton-Jacobi equation, where the velocity of the particle $v$ is given by
\begin{equation}
v=\frac{\nabla S}{m},
\end{equation}
which is known as the guidance equation and 
\begin{equation}
Q \equiv-\frac{\nabla^2 |\Psi|}{2m|\Psi|}
\end{equation}
is called the quantum potential. 
The equation of motion is
\begin{equation}
m\frac{d^2 x}{dt^2}=-\nabla V-\nabla Q.
\end{equation}
Note that this interpretation is deterministic. The wave function is a field (analogous to the electric field in electromagnetism) that gives a force to the particle via the quantum potential and guides its trajectory. The uncertainty of quantum mechanics becomes the unknown initial condition which is the hidden variable in this framework. A particle (such as an electron) guided by a wave function is often described as a drone being piloted by radio signals. 

It is certainly interesting to apply the de Broglie-Bohm mechanics to quantum cosmology. First of all, the measurement problem does not exist because there is no collapsing of the wave function and no requirement for an observer to do it. Also, the meaning of what is a singularity is more unambiguous in this context.

\section{minisuperspace Wheeler-DeWitt equation}
We consider a homogeneous and isotropic Friedmann-Lemaitre-Roberson-Walker (FLRW) metric 
\begin{equation}
ds^2=N^2dt^2-a^2d\Omega^2_k,
\end{equation}
as the minisuperspace mentioned in the introduction section. Here $N$ is the lapse function, $a\equiv e^\alpha$ is the scale factor, and $d\Omega^2_k$ is the spatial line element with curvature $k=-1,0,+1$. Note that $\dot{\alpha}$ is the Hubble parameter. In the following, we will choose the gauge $N=1$ (this makes $t$ so-called cosmic proper time), a flat metric $k=0$, and set $8\pi G=1$.
We consider a scalar field $\phi$ with potential $V$ as the matter field. If there is a cosmological constant, we combine it into the constant part of $V$. Since $\sqrt{-g}=e^{3\alpha}$, the classical Lagrangian is\footnote{The Ricci scalar is quadratic in terms of the extrinsic curvature and the extrinsic curvature in FLRW metric is proportional to the Hubble parameter $\dot{\alpha}$. We have used $k=0$ here.} 
\begin{equation}
L=e^{3\alpha}\left( \frac{\dot{\phi}^2}{2}-3\dot{\alpha}^2-V \right).
\end{equation}
The momenta conjugate to $\alpha$ and $\phi$ are 
\begin{equation}
\pi_\alpha=\frac{\partial L}{\partial \dot{\alpha}}=-6e^{3\alpha}\dot{\alpha},   \;\;\;   \pi_\phi=\frac{\partial L}{\partial \dot{\phi}}=e^{3\alpha}\dot{\phi}.
\end{equation}
The corresponding Hamiltonian is
\begin{equation}
H=e^{-3\alpha}\left( \frac{1}{2}\pi_\phi^2-\frac{1}{12}\pi_\alpha^2+V \right).
\end{equation}
After canonical quantization, $\pi_\alpha \rightarrow -i\partial/\partial \alpha$ and $\pi_\phi \rightarrow -i\partial/\partial \phi$ the Wheeler-DeWitt equation, $H\Psi(\alpha, \phi)=0$ for the wave function of the universe is given by
\begin{equation}
-\frac{1}{12}\frac{\partial^2}{\partial \alpha^2}\Psi+\frac{1}{2}\frac{\partial^2}{\partial \phi^2}\Psi+q\frac{\partial}{\partial \alpha}\Psi-e^{6 \alpha}V\Psi=0,
\label{wdw}
\end{equation}
where $q$ is any real constant due to semi-general factor ordering \cite{Hartle:1983ai}.
It is possible to find solutions for this equation for some particular potentials, such as $V=V_0e^{-2\sqrt{9+\mu^2 V_0}\Delta \phi}$ \cite{Guzman:2005xt}.
An exponential matter potential of the form $V=V_0e^{-\lambda \phi}$ is also considered for the study of bouncing universe to avoid big bang singularity \cite{Pinto-Neto:2018zvn, Colin:2017dwv, Bacalhau:2017hja}. Exact solutions of the Wheeler-DeWitt equation for a potential $V=\frac{1}{2}\mu(a) \phi^2+\frac{1}{4}\lambda(a)\phi^4$ with $a$-dependent coupling and a special non-FLRW minisuperspace is considered in \cite{Vink:1990fm}.
We would like to find a solution for simpler potential as a function of $\phi$ which ideally can be connected to potentials one may find in particle physics, such as a mass term in an FLRW universe.

\section{de Broglie-Bohm quantum cosmology}

We will focus on the de Broglie-Bohm interpretation of quantum cosmology \cite{Vink:1990fm} (see \cite{Pinto-Neto:2004szq, Pinto-Neto:2018zvn, Shtanov:1995ie, Goldstein:1999my, Pinto-Neto:2021jko} for review articles). Here the position of a particle considered in section \ref{sec2} is replaced by the field value $\phi$ and the parameter $\alpha$ which determines the scale factor via $a=e^\alpha$.
The time evolutions of $\phi$ and $\alpha$ are given by the guidance equations
\begin{equation}
\dot{\phi}=\frac{\partial_\phi S}{e^{3\alpha}},
\label{pdot}
\end{equation}
and 
\begin{equation}
\dot{\alpha}=-\frac{\partial_\alpha S}{6 e^{3\alpha}}.
\label{adot}
\end{equation}
Assuming $\Psi = |\Psi|e^{iS}$. We can make derivatives with respect to $\alpha$ and $\phi$ to obtain
\begin{eqnarray}
\frac{\partial\Psi}{\partial \alpha}&=&\partial_\alpha |\Psi|e^{iS}+|\Psi|e^{iS}i \partial_\alpha S    ,\\
\frac{\partial^2 \Psi}{\partial \alpha^2}&=&\partial^2_\alpha |\Psi|e^{iS}+2\partial_\alpha |\Psi|e^{iS}i\partial_\alpha S+|\Psi|e^{iS}(i\partial_\alpha S)^2+|\Psi|e^{iS}i\partial^2_\alpha S   ,\\
\frac{\partial^2 \Psi}{\partial \phi^2}&=&\partial^2_\phi |\Psi|e^{iS}+2\partial_\phi |\Psi|e^{iS}i\partial_\phi S+|\Psi|e^{iS}(i\partial_\phi S)^2+|\Psi|e^{iS}i\partial^2_\phi S.
\end{eqnarray}
Substituting these results into Eq.~(\ref{wdw}) and divided by $e^{6\alpha}\Psi$, the real part gives
\begin{equation}
\frac{1}{12}(\partial_\alpha S)^2-\frac{1}{2}(\partial_\phi S)^2-e^{6\alpha}(V+Q_M+6Q_G-q\frac{\partial_\alpha |\Psi|}{e^{6\alpha}|\Psi|})=0,
\label{s}
\end{equation}
where
\begin{equation}
Q_M \equiv -\frac{1}{2e^{6\alpha}}\frac{\partial^2_\phi |\Psi|}{|\Psi|}, \;\;\; Q_G \equiv \frac{1}{72e^{6\alpha}}\frac{\partial^2_\alpha |\Psi|}{|\Psi|}
\label{qp}
\end{equation}
are the quantum potentials.
By using Eqs.~(\ref{pdot}) and (\ref{adot}) to eliminate $S$, we obtain
\begin{equation}
3 \dot{\alpha}^2=\frac{\dot{\phi}^2}{2}+V+Q_M+6Q_G-q\frac{\partial_\alpha |\Psi|}{e^{6\alpha}|\Psi|}.
\label{m1}
\end{equation}
This equation is just the classical Friedmann equation with the last three terms as its "quantum corrections".
The equation of motion of $\phi$ can be obtained by taking derivative of Eq.~(\ref{s}) with respect to $\phi$ 
\begin{equation}
\ddot{\phi}+3\dot{\alpha}\dot{\phi}+\partial_\phi(V+Q_M+6Q_G-q\frac{\partial_\alpha |\Psi|}{e^{6\alpha}|\Psi|})=0,
\label{m2}
\end{equation}
where the relation $(\partial_\phi S\dot{)}=\dot{\phi}\partial^2_\phi S+\dot{\alpha}\partial_\phi \partial_\alpha S$ is used. This is just the classical Klein-Gorden equation in an expanding universe with the last three terms as "quantum corrections".

\section{The solution of an expanding and contracting flat universe}
\label{sec5}
\begin{figure}[t]
  \centering
\includegraphics[width=0.6\textwidth]{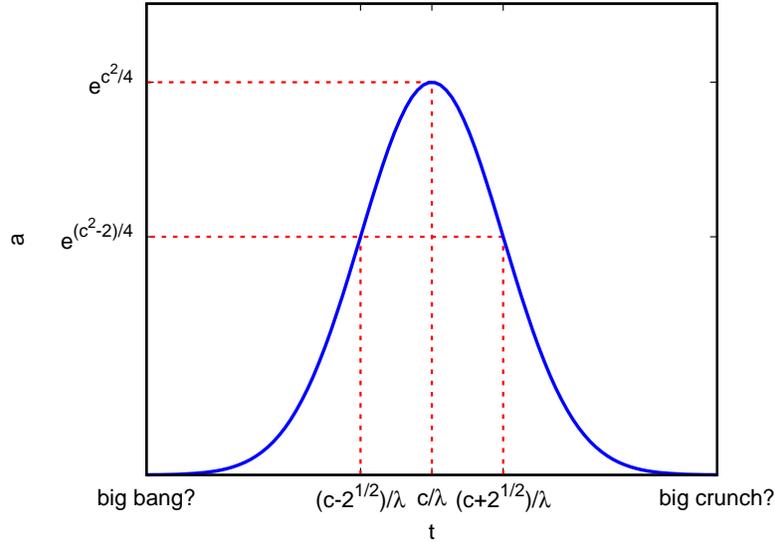}
  \caption{The scale factor $a$ as a function of time $t$. Different from the conventional big bang or big crunch, here the singularity at $a=0$ is never achieved.}
  \label{fig1}
\end{figure}

Let us consider an ansatz
\begin{equation}
\Psi=e^{iS}=e^{-i\lambda e^{3\alpha}\phi}.
\label{s1}
\end{equation}
Taking derivatives to obtain
\begin{eqnarray}
\frac{\partial^2 \Psi}{\partial \alpha^2}&=&-9\phi^2e^{6\alpha}\lambda^2e^{-i\lambda e^{3\alpha}\phi}-9\phi e^{3\alpha}\lambda e^{-i\lambda e^{3\alpha}\phi},\\
\frac{\partial^2 \Psi}{\partial \phi^2}&=&-\lambda^2 e^{6\alpha} e^{i\lambda e^{3\alpha}\phi}.
\end{eqnarray}
Substituting into the Wheeler-DeWitt equation of Eq.~(\ref{wdw}), 
\begin{equation}
\frac{3}{4}\phi^2 e^{6\alpha} \lambda^2 e^{-i\lambda e^{3\alpha}\phi}-\frac{\lambda^2}{2}e^{6\alpha} e^{-i\lambda e^{3\alpha}\phi}-3iq\lambda \phi e^{3\alpha} e^{-i\lambda e^{3\alpha}\phi}+\frac{3i}{4}\lambda \phi e^{3 \alpha} e^{-i\lambda e^{3\alpha}\phi}-e^{6\alpha}V e^{-i\lambda e^{3\alpha}\phi}=0.
\end{equation}
If we choose $q=\frac{1}{4}$, the imaginary terms cancelled, and we are left with
\begin{equation}
V= \lambda^2\left( \frac{3}{4}\phi^2-\frac{1}{2} \right) \equiv V_0 \left( \frac{3}{4}\phi^2-\frac{1}{2} \right).
\label{p1}
\end{equation}
This is a quadratic potential, namely, a mass term of the scalar field plus a negative cosmological constant. 
Therefore Eq.~(\ref{s1}) is an exact solution, where
\begin{equation}
S=-\lambda e^{3\alpha}\phi.
\end{equation}
From the guidance equation of Eq.~(\ref{pdot}), we have
\begin{equation}
\dot{\phi}=\frac{\partial_\phi S}{e^{3\alpha}}=-\lambda.
\end{equation}
This means the scalar field $\phi$ is rolling down the potential given by Eq.~(\ref{p1}) at a constant speed. It can be integrated to
\begin{equation}
\phi=-\lambda t +c,
\label{phi}
\end{equation}
where $c$ is an integration constant.
From the guidance equation of Eq.~(\ref{adot}), we obtain
\begin{equation}
\dot{\alpha}=-\frac{\partial_\alpha S}{6e^{3\alpha}}=\frac{\lambda\phi}{2}=-\frac{\lambda^2}{2}t+\frac{\lambda}{2}c.
\end{equation}
This can be integrated to
\begin{equation}
\alpha=-\frac{\lambda^2}{4}t^2+\frac{\lambda}{2}ct.
\end{equation}
The scale factor is given by
\begin{equation}
a = e^\alpha=e^{-\frac{\lambda^2}{4}\left(t-\frac{c}{\lambda}\right)^2+\frac{c^2}{4}}.
\end{equation}
This is a Gaussian function. To my knowledge, this is for the first time a Gaussian function of the evolution of the scale factor is obtained. It describes a universe that begins at $a \rightarrow 0$ at $t \rightarrow -\infty$ (the big bang?). It underwent an accelerating inflationary phase until $t=\frac{c-\sqrt{2}}{\lambda}$ and the scale factor starts a decelerating phase. The scale factor reaches a maximum value at $t=\frac{c}{\lambda}$ (which corresponds to $\phi=0$) and starts to decrease. After $t=\frac{c+\sqrt{2}}{\lambda}$ has been achieved, the scale factor continues decreasing but accelerating\footnote{Usually inflation is simply defined as the period when $\ddot{a}>0$. Here we have $\ddot{a}>0$ but $\dot{a}<0$. Probably people would rather change the definition this time and prefer not to call it inflation.}. Eventually, we have $a \rightarrow 0$ at $t \rightarrow \infty$ (the big crunch?). Note that we are considering a flat universe, not a closed one. For clarity, we plot this result in Fig.~\ref{fig1}.

Throughout the evolution of this universe, the scalar field is rolling at a constant speed and there is no singularity since $a=0$ cannot be achieved. If someone insisted that we should start from $t=0$, we can define the field value at $t=0$ as $\phi_0$ and $c = \phi_0$ from Eq.~(\ref{phi}). If someone insisted that the energy density of the scalar field (namely, $\frac{\dot{\phi}^2}{2}+V$) should not exceed the Planck energy, a maximum field value $\bar{\phi}=2/\sqrt{3}\lambda$ can be obtained. Someone may go further and suggest $\phi_0=\bar{\phi}$, then $a \rightarrow 0$ will not happen. It turns out the scale factor will be $a=1$ at $t=0$. Well, the value does not mean anything because only the ratio of the scale factor matters. 


From Eq.~(\ref{qp}), we can see that $Q_M=Q_G=\partial_\alpha |\Psi|=0$ since $|\Psi|=1$. Therefore this solution coincides with the classical solution of the evolution of a scalar field $\phi$ with its potential given by Eq.~(\ref{p1}). Let us check it. In this case, Eq.~(\ref{m1}) becomes
\begin{equation}
3\dot{\alpha}^2=\frac{\dot{\phi}^2}{2}+V=\frac{\dot{\phi}^2}{2}+\frac{3\lambda^2 \phi^2}{4}-\frac{\lambda^2}{2}.
\end{equation}
On the other hand, Eq.~(\ref{m2}) becomes
\begin{equation}
\ddot{\phi}+3\dot{\alpha}\dot{\phi}+\partial_\phi V=\ddot{\phi}+3\dot{\alpha}\dot{\phi}+\frac{3\lambda^2 \phi}{2}=0.
\end{equation}
It is easy to see that $\dot{\phi}=-\lambda$ and $\dot{\alpha}=\lambda \phi/2$ provide a solution to the above equations.
Note that usually when the equation of motion of a homogeneous scalar field in an expanding universe is considered, some approximation scheme is considered, such as slow-roll approximation. However, for this particular potential, there is an exact solution without approximation.

Although we focus on the de Broglie-Bohm interpretation. The solution of the Wheeler-DeWitt equation given in Eq.~(\ref{s1}) is also valid for any other interpretation. Some may interpret that the wave function is peaked about solutions to the classical Einstein equations. It is intriguing to contemplate the physical meaning of $|\Psi|^2=1$ here. 

\section{The solution of a Minkowski universe}
One may feel it is not very interesting if the quantum corrections vanish as $Q_M=Q_G=\partial_\alpha |\Psi|=0$. There is another solution that can be obtained by simply changing $\lambda \rightarrow i\lambda$ in Eq.~(\ref{s1}).
\begin{equation}
\Psi=|\Psi|=e^{\lambda e^{3\alpha}\phi}.
\label{s2}
\end{equation}
Substituting into the Wheeler DeWitt equation of Eq.~(\ref{wdw}), we obtain 
\begin{equation}
-\frac{3}{4}\phi^2 e^{6\alpha} \lambda^2 e^{\lambda e^{3\alpha}\phi}+\frac{\lambda^2}{2}e^{6\alpha} e^{\lambda e^{3\alpha}\phi}+3q\lambda \phi e^{3\alpha} e^{\lambda e^{3\alpha}\phi}-\frac{3}{4}\lambda \phi e^{3 \alpha} e^{\lambda e^{3\alpha}\phi}-e^{6\alpha}V e^{\lambda e^{3\alpha}\phi}=0.
\end{equation}
Again by setting $q=1/4$, we have
\begin{equation}
V= \lambda^2\left(\frac{1}{2}-\frac{3}{4}\phi^2 \right) \equiv V_0 \left( \frac{1}{2}-\frac{3}{4}\phi^2 \right).
\label{p2}
\end{equation}
This potential is of a hilltop form. Although it seems to be unbounded from below, we can introduce a quartic term $\sim \phi^4$ to stabilize the potential if necessary. When the field value of $\phi$ is small enough, namely close to the hilltop, the quartic term can be neglected. With a suitable choice of $V_0$, the field $\phi$ can even be the Higgs field in the standard model of particle physics! It can also be a tachyon. Let us see what lesson we can learn from this solution. Here we have a real solution, therefore $S=0$ in the phase. From Eqs.~(\ref{pdot}) and (\ref{adot}), the guidance equations gives $\dot{\phi}=\dot{\alpha}=0$. This can be integrated into the constants $\phi=c_1$ and $\alpha=c_2$. The field $\phi$ and the scale factor $a=e^\alpha$ are "frozen". This describes a static Minkowski universe. But how come $\phi$ does not roll at all with the potential given by Eq.~(\ref{p2})? From Eq.~(\ref{qp}), we obtain the quantum potentials
\begin{equation}
Q_M=-\frac{\lambda^2}{2}, \;\;\;  Q_G=\frac{\lambda^2 \phi^2}{8}+\frac{\phi}{8e^{3\alpha}}.
\end{equation}
Therefore 
\begin{equation}
Q_M+6Q_G-q\frac{\partial_\alpha |\Psi|}{e^{6\alpha}|\Psi|}+V=-\frac{\lambda^2}{2}+\frac{3\lambda^2 \phi^2}{4}+V=0.
\end{equation}
 In this case, Eq.~(\ref{m1}) becomes
\begin{equation}
3\dot{\alpha}^2=\frac{\dot{\phi}^2}{2},
\end{equation}
and Eq.~(\ref{m2}) becomes
\begin{equation}
\ddot{\phi}+3\dot{\alpha}\dot{\phi}=0.
\end{equation}
It is easy to see that constant $\phi$ and $\alpha$ provide a solution to the above equations. The quantum effects exactly canceled out the classical potential and produce a Minkowski universe. We somehow find a nontrivial way to produce a Minkowski universe with non-vanishing classical and quantum potentials. For a constant $\phi$, the potential behaves like a cosmological constant. It seems we can have a Minkowski universe with an arbitrarily large negative cosmological constant. 
This is opposite to the original motivation of Einstein to introduce a positive cosmological constant to obtain a static universe! This may shed some light on the meaning of dark energy or zero-point energy.

Last but not least, we would like to mention that different from $|\Psi|^2=1$ in section \ref{sec5}, here we have $|\Psi|^2=e^{2\lambda e^{3\alpha}\phi}$. It may be interesting to think about whether there is a probability interpretation for this result.
\section{conclusion}
In this work, we have presented two simple solutions for the Wheeler DeWitt equation in a flat minisuperspace universe. The universe may be expanding, contracting, or static.
The evolution of the universe described here may not be realistic given the simple wave function. However, at least we have a solution to play with and we wish it would provide a framework to try to find more solutions. Although we focus on de Broglie-Bohm interpretation, our solution of the Wheeler-DeWitt equation can be applied to any other interpretation. It is useful when we try to think about what is the physical meaning of the wave function of the universe. We hope the reader of this article can test their favorite interpretation by using the exact solutions as a tool. 

As is shown in this work, quantum cosmology is not necessarily a theory about the beginning of the universe. It can play some role even in a static universe. The wave function of the universe may always be there guiding the evolution of the matter field and the scale factor throughout the history of the universe.

\acknowledgments
This work is supported by the National Science and Technology Council (NSTC) of Taiwan under Grant No. NSTC 111-2112-M-167-002.


\begin{thebibliography}{99}

\bibitem{DeWitt:1967yk}
B.~S.~DeWitt,
Phys. Rev. \textbf{160}, 1113-1148 (1967)
doi:10.1103/PhysRev.160.1113

\bibitem{Halliwell:1989myn}
J.~J.~Halliwell,
[arXiv:0909.2566 [gr-qc]].

\bibitem{Bojowald:2015iga}
M.~Bojowald,
Rept. Prog. Phys. \textbf{78}, 023901 (2015)
doi:10.1088/0034-4885/78/2/023901
[arXiv:1501.04899 [gr-qc]].

\bibitem{Linde:1990flp}
A.~D.~Linde,
Contemp. Concepts Phys. \textbf{5}, 1-362 (1990)
[arXiv:hep-th/0503203 [hep-th]].

\bibitem{Kolb:1990vq}
E.~W.~Kolb and M.~S.~Turner,
Front. Phys. \textbf{69}, 1-547 (1990)
doi:10.1201/9780429492860

\bibitem{Tryon:1973xi}
E.~P.~Tryon,
Nature \textbf{246}, 396 (1973)
doi:10.1038/246396a0

\bibitem{Bohr:1935af}
N.~Bohr,
Phys. Rev. \textbf{48}, 696-702 (1935)
doi:10.1103/PhysRev.48.696

\bibitem{Guth:1980zm}
A.~H.~Guth,
Phys. Rev. D \textbf{23}, 347-356 (1981)
doi:10.1103/PhysRevD.23.347

\bibitem{Bohm:1951xw}
D.~Bohm,
Phys. Rev. \textbf{85}, 166-179 (1952)
doi:10.1103/PhysRev.85.166


\bibitem{Hartle:1983ai}
J.~B.~Hartle and S.~W.~Hawking,
Phys. Rev. D \textbf{28}, 2960-2975 (1983)
doi:10.1103/PhysRevD.28.2960

\bibitem{Guzman:2005xt}
W.~Guzman, M.~Sabido, J.~Socorro and L.~Arturo Urena-Lopez,
Int. J. Mod. Phys. D \textbf{16}, 641-654 (2007)
doi:10.1142/S0218271807009401
[arXiv:gr-qc/0506041 [gr-qc]].


\bibitem{Pinto-Neto:2018zvn}
N.~Pinto-Neto and W.~Struyve,
[arXiv:1801.03353 [gr-qc]].

\bibitem{Colin:2017dwv}
S.~Colin and N.~Pinto-Neto,
Phys. Rev. D \textbf{96}, no.6, 063502 (2017)
doi:10.1103/PhysRevD.96.063502
[arXiv:1706.03037 [gr-qc]].

\bibitem{Bacalhau:2017hja}
A.~P.~Bacalhau, N.~Pinto-Neto and S.~Dias Pinto Vitenti,
Phys. Rev. D \textbf{97}, no.8, 083517 (2018)
doi:10.1103/PhysRevD.97.083517
[arXiv:1706.08830 [gr-qc]].

\bibitem{Vink:1990fm}
J.~C.~Vink,
Nucl. Phys. B \textbf{369}, 707-728 (1992)
doi:10.1016/0550-3213(92)90283-H

\bibitem{Pinto-Neto:2004szq}
N.~Pinto-Neto,
Found. Phys. \textbf{35}, 577-603 (2005)
doi:10.1007/s10701-004-2012-8
[arXiv:gr-qc/0410117 [gr-qc]].

\bibitem{Shtanov:1995ie}
Y.~Shtanov,
Phys. Rev. D \textbf{54}, 2564-2570 (1996)
doi:10.1103/PhysRevD.54.2564
[arXiv:gr-qc/9503005 [gr-qc]].

\bibitem{Goldstein:1999my}
S.~Goldstein and S.~Teufel,
[arXiv:quant-ph/9902018 [quant-ph]].

\bibitem{Pinto-Neto:2021jko}
N.~Pinto-Neto,
Universe \textbf{7}, no.5, 134 (2021)
doi:10.3390/universe7050134
[arXiv:2111.03057 [gr-qc]].


\end{thebibliography}
\end{document}